\documentclass[11pt]{article}

\usepackage[margin=1in]{geometry}
\usepackage[T1]{fontenc}
\usepackage[utf8]{inputenc}
\usepackage{newtxtext,newtxmath} 
\usepackage[cal=cm,bb=ams]{mathalfa} 
\usepackage{microtype}

\usepackage{authblk}
\usepackage{abstract}
\usepackage{fancyhdr}
\usepackage{titlesec}

\usepackage{amsmath}
\usepackage{amsfonts}
\usepackage{mathtools}
\usepackage{bm}
\usepackage{physics}    
\usepackage{bbm}        

\usepackage{graphicx}
\usepackage{subcaption}
\usepackage{booktabs}
\usepackage{multirow}
\usepackage{makecell}

\usepackage[authoryear,round]{natbib}
\usepackage[
    colorlinks=true,
    citecolor=blue,
    linkcolor=blue,
    urlcolor=blue
]{hyperref}

\usepackage[nameinlink,capitalise,noabbrev]{cleveref}

\titleformat{\section}
  {\large\bfseries}
  {\thesection}
  {0.75em}
  {}

\titleformat{\subsection}
  {\normalsize\bfseries}
  {\thesubsection}
  {0.75em}
  {}

\title{
       \Large\bfseries
        A First Application of Split Conformal Prediction for Dose Prediction in Knowledge-Based Radiotherapy Planning
       }

\author[1, 2]{Raphaël Brodeur}
\author[3]{Anne-Sophie Charest}
\author[1, 2]{Louis Archambault}

\affil[1]{Département de physique, de génie physique et d'optique, Faculté des sciences et de génie, Université~Laval,~Québec,~Canada}
\affil[2]{Axe oncologie, Centre de recherche du CHU de Québec--Université Laval,
Québec, Canada}
\affil[3]{Département de mathématiques et de statistique, Université Laval, Québec, Canada}

\date{November 2026}

\begin{document}

\maketitle

\vspace{-1em}

{\small
\noindent\textbf{Corresponding author:} 
louis.archambault@phy.ulaval.ca

\noindent\textbf{Keywords:}
Deep learning;
Dose prediction intervals;
Radiotherapy treatment planning;
Split conformal prediction;
Uncertainty quantification
}

\vspace{0.5em}

\noindent\rule{\textwidth}{0.8pt}

\begin{abstract}
\noindent
\textit{Objective.}
In radiation therapy, artificial intelligence (AI) has emerged as a promising avenue for automating and standardising treatment planning,
a labour-intensive process prone to inter-expert variability.
However, most deep learning (DL) dose prediction models provide point predictions without explicitly quantifying their reliability;
this is a key concern for their use in safety-critical clinical applications.
In this study, we construct prediction intervals for DL-based radiotherapy dose prediction using split conformal prediction (CP),
a model-agnostic and distribution-free method for uncertainty quantification that provides finite-sample coverage guarantees.
\textit{Approach.}
We propose patient-level nonconformity scores obtained by aggregating voxelwise scores.
Specifically, we formulate
a maximum aggregation construction, which guarantees simultaneous coverage of the true dose in all voxels of a future patient with at least a user-defined probability,
and a quantile-based construction, which guarantees, with the same probability lower bound, that a future patient attains a prescribed minimum level of voxel coverage.
We also investigate groupwise calibration for region-specific guarantees
and predictive uncertainty scaling for locally adaptive prediction intervals.
The methods are evaluated on head-and-neck cases using a deep ensemble model.
\textit{Main results.}
The proposed constructions achieved empirical coverage consistent with the nominal levels.
Maximum aggregation led to highly conservative prediction intervals,
whereas quantile aggregation produced substantially narrower intervals while targeting a less stringent coverage event.
Groupwise calibration and uncertainty scaling reduced interval widths in some anatomical regions,
while providing region-specific guarantees and local adaptivity, respectively.
\textit{Significance.}
To our knowledge, this is the first application of split CP to radiotherapy dose prediction.
The explicit patient-level coverage guarantees provided by the proposed split CP constructions offer a more interpretable characterisation of prediction reliability than raw uncertainty estimates.
In a potential clinical application of AI-based dose prediction, split CP could help clinicians assess the reliability of model predictions and identify cases requiring further review.
\end{abstract}

\vspace{-0.5em}
\noindent\rule{\textwidth}{0.8pt}

\section{Introduction}
\label{sec:introduction}

Recently, deep learning (DL) dose prediction models have shown promising performance for knowledge-based treatment planning,
with the potential to accelerate the planning process and improve standardisation~\citep{nguyen-2019, lagedamon-2024}.
However, most models produce point predictions of the 3D dose distribution without communicating how reliable a predicted dose distribution is for a given patient.
In safety-critical applications such as radiotherapy,
uncertainty quantification (UQ) is essential for clinical integration of artificial intelligence (AI), as it enables clinicians to assess the extent to which model outputs may be relied upon and to flag cases requiring further review.

Several methods have been proposed for UQ in deep learning, such as Bayesian neural networks (BNNs) and ensemble methods~\citep{jospin-2022}.
In practice, these methods typically produce an uncertainty estimate alongside the model prediction, often by measuring the variability across multiple predictions for the same input.
This variability can be obtained in different ways depending on the UQ method.
Variational BNNs seek to approximate the posterior distribution over the network weights and generate stochastic predictions by sampling weight values from this approximation at inference time~\citep{graves-2011, blundell-2015}.
Monte Carlo dropout (MCD) interprets training with dropout as approximate variational inference and estimates predictive uncertainty by retaining dropout at test time and performing multiple stochastic forward passes~\citep{gal-2016}.
Deep ensembles estimate uncertainty from the variability among the predictions produced by multiple independently trained instances of the same model architecture~\citep{lakshminarayanan-2017}.
Other approaches include Laplace approximations~\citep{ritter-2018, daxberger-2021}, Gaussian stochastic weight averaging (SWAG)~\citep{maddox-2019}, and test-time augmentation (TTA)~\citep{ayhan-2018}, among others.

Although UQ is receiving increased attention in DL, its application to radiotherapy dose prediction remains limited~\citep{wahid-2024, tan-2024}.
Nevertheless, recent works have explored various methods to quantify uncertainty in dose prediction tasks.
These have focused on producing uncertainty estimates alongside model predictions,
using methods such as
MCD~\citep{nguyen-2021, tan-2024, huet-2024},
bagging~\citep{nguyen-2021},
deep ensembles~\citep{tan-2024, huet-2024},
deep evidential learning~\citep{tan-2024},
mixture density networks~\citep{nilsson-2021},
and input reconstruction error as a surrogate for model uncertainty~\citep{huet-2024}.

Existing UQ approaches for DL can offer useful uncertainty estimates,
but these estimates do not, by themselves, constitute a guarantee on prediction errors.
Furthermore, while informative, commonly used uncertainty measures such as predictive variance and entropy may be difficult to interpret for an end user in a clinical setting,
and especially across institutions that may have different treatment planning practices.
By contrast, prediction intervals associated with an explicit guarantee on the probability of error can provide a more directly interpretable characterisation of predictive uncertainty.
More broadly, a scoping review by~\citet{wahid-2024} of UQ methods in various radiotherapy applications identified a clear need for studies on additional UQ methods, such as conformal prediction. 

Conformal prediction (CP)~\citep{papadopoulos-2002, vovk-2005, lei-2018} is a post hoc calibration framework for constructing prediction intervals equipped with an explicit guarantee on the probability that they contain the true target.
Specifically, it provides finite-sample coverage guarantees for future observations,
meaning that the guarantees hold for a finite calibration set rather than only asymptotically as the number of calibration samples tends to infinity. 
Moreover, CP is distribution-free and model-agnostic:
the validity of the coverage guarantees does not depend on assumptions about the underlying data-generating distribution or about the predictive model.
Instead, it relies primarily on exchangeability between calibration data and future observations, a comparatively mild assumption.
These properties make CP particularly relevant for UQ in DL, where predictive models are often complex and difficult to calibrate probabilistically.
The form of CP considered in this work is split CP~\citep{papadopoulos-2002}, in which a held out calibration set is used to construct prediction sets for future data.
Despite these advantages, to our knowledge, split CP has not yet been applied to dose prediction for radiotherapy treatment planning.

Extending CP to DL-based dose prediction, and more generally to image-to-image prediction tasks, is not straightforward.
Ideally, one would like to construct a prediction interval for the dose at each voxel of the output dose map, with each interval satisfying a valid conformal coverage guarantee.
However, individual voxels are not exchangeable observations;
rather, it is at the patient level that exchangeability can naturally be assumed.
This mismatch between the level at which exchangeability holds (patients) and the level at which prediction intervals are desired (voxels) poses a challenge.
In typical CP applications, each exchangeable example is associated with a single target for which coverage is sought,
whereas in dose prediction each exchangeable patient is associated with a large number of voxel values for which separate prediction intervals are desired.
The challenge is therefore to define nonconformity scores at the patient level in a way that allows prediction intervals to be constructed at the voxel level.

Closely related distribution-free approaches include
risk-controlling prediction sets (RCPS),
which guarantee with a prescribed probability over the calibration data that the expected value of a user-defined loss on future observations does not exceed a prescribed level~\citep{bates-2021},
and conformal risk control (CRC),
which guarantees that the expected value of a user-defined loss, marginally over the calibration and future observations, does not exceed a prescribed level~\citep{angelopoulos-crc-2024}.
\citet{angelopoulos-rcps-2022} adapted RCPS to image-to-image regression by controlling the expected fraction of miscovered pixels and introduced several approaches for constructing adaptive intervals, including pixelwise quantile regression.
\citet{fisher-2024} subsequently applied RCPS to DL-based radiotherapy dose calculation, calibrating voxelwise quantile regression intervals to control the expected voxel miscoverage risk and simultaneously controlling this risk across multiple spatial subgroups.
In contrast,
rather than controlling the expected voxel miscoverage risk,
our patient-level split CP construction provides a guarantee on the probability that a future patient attains at least a prescribed minimum fraction of voxel coverage.
Beyond regression,
similar coverage guarantees have also been obtained in image segmentation.
In particular,
\citet{mossina-2024} used CRC with binary losses to control image-level coverage events for pixelwise label prediction sets,
while~\citet{mossina-friedrich-2025} used split CP to control the coverage of segmentation masks expanded by morphological dilation.

In this work, we investigate the application of split CP to DL-based radiotherapy dose prediction for knowledge-based treatment planning,
in which past treatment plans are used to predict achievable dose distributions for new patients~\citep{appenzoller-2012}.
In particular, we address the mismatch between the patient level, at which exchangeability is assumed, and the voxel level, at which prediction intervals are desired, by formulating patient-level nonconformity scores that yield conformal prediction intervals for individual voxels.
We propose a patient-level nonconformity score based on maximum aggregation that provides a simultaneous conformal coverage guarantee over all voxels of a future patient,
and further derive from this guarantee a marginal coverage lower bound for every individual voxel.
We also propose a quantile-based patient-level score that guarantees that a future patient attains at least a prescribed minimum fraction of voxel coverage.
To improve the efficiency and adaptivity of the resulting intervals,
we further investigate two established approaches:
groupwise calibration to obtain region-specific guarantees,
and scaling by predictive uncertainty estimates to construct  spatially adaptive voxelwise prediction intervals.
Finally,
we analyse the implications and limitations of using voxel coverage as the target of conformal guarantees for dose prediction, particularly in relation to the width and practical usefulness of the resulting prediction intervals.

The remainder of this paper is organised as follows.
Section~\ref{sec:methods} describes the proposed CP methods, as well as the experiments and corresponding evaluation protocol.
Section~\ref{sec:results} presents the experimental results,
which are discussed in Section~\ref{sec:discussion}.
The main conclusions are then summarised in Section~\ref{sec:conclusion}.

\section{Methods}
\label{sec:methods}

\textit{Notation.}
Throughout this work, bold symbols denote vectors and matrices, whereas non-bold symbols denote scalars.
Capital and lowercase letters denote random variables and their realisations, respectively.
Superscripts in parentheses index samples in a dataset.
Depending on the context, subscripts may index stochastic predictions,
indicate a parametrisation,
or specify an element of an object (e.g. $a_i$ is the $i$-th element of vector $\vb{a}$).

\subsection{General framework of split conformal prediction}

The finite-sample coverage guarantee of split CP relies on exchangeability between the calibration examples and a future test example.
A collection of random variables is said to be exchangeable if its joint distribution is invariant to permutations of their ordering.
More precisely, CP outputs for a future test input $X^{(\text{test})} \in \mathcal{X}$ a prediction set $\mathcal{C}(X^{(\text{test})}) \subseteq \mathcal{Y}$ containing plausible target values.
Crucially, under exchangeability of the $n_{\text{cal}}$ data points held out for calibration and a future test pair $(X^{(\text{test})}, Y^{(\text{test})}) \in \mathcal{X} \times \mathcal{Y}$, the following marginal coverage guarantee holds~\citep{lei-2018}:
\begin{equation}
\label{eq:marginal-coverage-guarantee}
1 - \alpha
\leq
\mathbb{P}\left[ 
Y^{(\text{test})} \in \mathcal{C}(X^{(\text{test})})
\right]
\leq
1 - \alpha + \frac{1}{n_{\text{cal}} + 1},
\end{equation}
where $1-\alpha \in (0, 1)$ is a user-specified desired coverage level,
and the probability is taken over the true joint distribution of the $n_{\text{cal}} + 1$ draws forming the calibration set and the future test example.
In words, the probability that the prediction set contains the true target value is at least, and almost exactly\footnote{The upper bound further demands a no-ties assumption on the $n_\text{cal}+1$ nonconformity scores.}, the desired coverage level $1-\alpha$, marginally over the calibration and test data~\citep{angelopoulos-2023}.

The split CP procedure is as follows.
First, consider a trained predictive model $f_{\hat{\boldsymbol{\theta}}} \colon \mathcal{X} \rightarrow \mathcal{Y}$ and a calibration set $\{ (X^{(i)}, Y^{(i)})\}_{i=1}^{n_{\text{cal}}}$ composed of $n_{\text{cal}}$ examples unseen during training.
A nonconformity function $s\colon \mathcal{X} \times \mathcal{Y} \rightarrow \mathbb{R}_{\geq 0}$ is then defined
such that the nonconformity score $s(x, y)$ quantifies the degree to which a candidate target value $y$ is incompatible with the input $x$.
In practice, the nonconformity function is typically defined using $f_{\hat{\boldsymbol{\theta}}}$ so that the resulting score measures the disagreement between a candidate target value and the corresponding model prediction.
The nonconformity function is then evaluated on each calibration example, yielding calibration scores
$s^{(i)} \coloneq s(X^{(i)}, Y^{(i)})$ for $i=1, \ldots, n_{\text{cal}}$.
For a target coverage level $1-\alpha \in (0, 1)$, the conformal threshold $\hat{q}$ is then defined as an empirical quantile of the calibration scores:
\begin{equation}
\label{eq:q-hat}
\hat{q} \coloneq
\lceil (n_{\text{cal}} + 1)(1 - \alpha) \rceil
\text{-th smallest score in }
\left\{ s^{(i)} \right\}_{i=1}^{n_\text{cal}}.
\end{equation}
This step is referred to as calibration.
Finally, for a new test input $X^{(\text{test})}$, the conformal prediction set is constructed using the calibrated threshold $\hat{q}$:
\begin{equation}
\label{eq:conformal-prediction-set}
\mathcal{C}(X^{(\text{test})}) \coloneq
\left\{
y \in \mathcal{Y} :
s(X^{\text{(test)}}, y) \leq \hat{q}
\right\}.
\end{equation}
That is, the conformal prediction set contains all possible target values whose nonconformity score with respect to the test input does not exceed the conformal threshold.
Under exchangeability, this construction yields the marginal coverage guarantee stated in~\eqref{eq:marginal-coverage-guarantee}.

\subsection{Limitations of voxel-level calibration approaches}

For scalar regression, a common choice of nonconformity score is the absolute residual between the target value and the model prediction.
Extending this idea to dose prediction, for an input $\vb{x}$ and corresponding ground truth dose map $\vb{y}=[y_v]_{v=1}^V$, we define the nonconformity score at voxel $v$ as
\begin{equation}
\label{eq:voxelwise-residual}
s_v(\vb{x}, y_v) \coloneq \abs{y_v - f_{\hat{\boldsymbol{\theta}}}(\vb{x})_v}.
\end{equation}
A seemingly straightforward calibration strategy,
which we refer to as pooled voxel-level calibration,
is then to pool these voxelwise scores across all calibration patients and voxels and calibrate a single threshold $\hat{q}$.
Substituting this score into the prediction set construction in~\eqref{eq:conformal-prediction-set} would yield the symmetric prediction interval
$[f_{\hat{\boldsymbol{\theta}}}(\vb{x})_v - \hat{q}, f_{\hat{\boldsymbol{\theta}}}(\vb{x})_v + \hat{q}]$ for each voxel $v$.
However, pooling the scores in this way implicitly treats individual voxel scores as exchangeable calibration observations, whereas exchangeability is naturally assumed at the patient level.
Consequently, the split CP marginal coverage guarantee does not apply to the resulting intervals.

An alternative is to perform calibration voxel-by-voxel by computing a distinct threshold $\hat{q}_v$ for each voxel position from the scores obtained at the same voxel index across the calibration patients.
Under patient exchangeability, this yields a separate marginal coverage guarantee for each voxel index $v$.
However, the same voxel index does not necessarily correspond to the same anatomical location across patients;
for example, the same voxel index may lie within the head for one patient and outside the body for another.
While valid, voxelwise calibration is therefore poorly suited to dose prediction.
These limitations motivate defining nonconformity scores at the patient level, in a way that nevertheless yields prediction intervals at the voxel level.

\subsection{Patient-level nonconformity scores}

Because exchangeability is assumed at the patient level, we instead define a single scalar nonconformity score for each patient by aggregating the voxelwise nonconformity scores over the dose map.
Calibration is then performed on these patient-level nonconformity scores, with each calibration patient contributing one score, 
thus allowing for the split CP coverage guarantee to apply.
The choice of aggregation is critical, as it defines the event whose probability is controlled by CP and, consequently, the interpretation of the resulting prediction intervals.
In the following, we propose two patient-level nonconformity scores based respectively on maximum and quantile aggregation of the voxelwise nonconformity scores.

\paragraph{Maximum aggregation.}

We first propose a patient-level nonconformity score based on the maximum of the voxelwise nonconformity scores.
More precisely, for a patient with input $\vb{x}$ and ground truth dose map $\vb{y}$, we define the nonconformity score as
\begin{equation}
\label{eq:patient-level-score-max}
s_{\text{max}}(\vb{x}, \vb{y}) \coloneq \max_v \abs{y_v - f_{\hat{\boldsymbol{\theta}}}(\vb{x})_v}.
\end{equation}
Each calibration patient therefore contributes its largest voxelwise nonconformity score as a single scalar patient-level nonconformity score.
After computing $\hat{q}$ from the patient-level calibration scores using~\eqref{eq:q-hat}, the resulting conformal prediction set for an input $\vb{x}$ is
\begin{equation}
\label{eq:conformal-prediction-set-max}
\mathcal{C}_{\text{max}}(\vb{x}) =
\left\{
\vb{y} \in \mathcal{Y} :
\forall \, v, \; \abs{y_v - f_{\hat{\boldsymbol{\theta}}}(\vb{x})_v} \leq \hat{q}
\right\}.
\end{equation}
Equivalently, the conformal prediction set $\mathcal{C}_{\text{max}}(\vb{x})$ contains all candidate dose maps $\vb{y}$ whose dose $y_v$ at every voxel $v$ lies within the corresponding voxelwise prediction interval
\begin{equation}
\mathcal{I}_v(\vb{x}) \coloneq
[f_{\hat{\boldsymbol{\theta}}}(\vb{x})_v -\hat{q}, \,
f_{\hat{\boldsymbol{\theta}}}(\vb{x})_v +\hat{q}].
\end{equation}
Under exchangeability of the calibration data and a future test pair $(\vb{X}^{(\text{test})}, \vb{Y}^{(\text{test})})$, it follows from~\eqref{eq:marginal-coverage-guarantee} that this construction satisfies
\begin{equation}
\label{eq:max-nc-score-guarantee}
\mathbb{P}\left[ 
\forall \, v, \; Y_v^{(\text{test})} \in \mathcal{I}_v(\vb{X}^{(\text{test})})
\right] \geq 1 - \alpha.
\end{equation}
That is, this construction guarantees with probability at least $1-\alpha$ in the sense of~\eqref{eq:marginal-coverage-guarantee} that the voxelwise prediction intervals simultaneously contain the true dose at every voxel of a future patient.

Analogous image-level coverage guarantees have previously been considered in image segmentation.
In particular, \citet{mossina-2024} used CRC with a binary loss corresponding to simultaneous coverage of ground truth labels by the pixelwise prediction sets,
while~\citet{mossina-friedrich-2025} used split CP with morphological dilation of the predicted segmentation mask to construct a prediction set that can target complete coverage of the ground truth mask.
Although these approaches target the same coverage event as our maximum aggregation construction,
namely simultaneous coverage,
and provide guarantees analogous to~\eqref{eq:max-nc-score-guarantee} in the image segmentation setting,
their constructions differ substantially.
Beyond the difference in the nature of the task (classification as opposed to regression),
\citet{mossina-2024} used CRC with a binary image-level loss,
while~\citet{mossina-friedrich-2025} used split CP with a nonconformity score based on the number of morphological dilations required for the dilated predicted mask to cover a prescribed fraction of the ground truth mask;
in contrast, our construction uses split CP with a patient-level nonconformity score obtained by maximum aggregation of voxelwise scores,
yielding voxelwise continuous dose prediction intervals.

The simultaneous coverage guarantee in~\eqref{eq:max-nc-score-guarantee} further implies a marginal coverage guarantee for every individual voxel.
Indeed, for any specific voxel $v$, the event that all voxelwise intervals cover their respective true doses is contained in the event that the interval at voxel $v$ covers its respective true dose $Y_v^{(\text{test})}$.
By monotonicity of probability, it thus follows from~\eqref{eq:max-nc-score-guarantee} that, for any fixed voxel $v$,
\begin{equation}
\label{eq:max-score-induced-voxelwise-guarantee}
\mathbb{P}\left[
Y_v^{(\text{test})} \in \mathcal{I}_v(\vb{X}^{(\text{test})})
\right] \geq 1 - \alpha.
\end{equation}
Therefore, although calibration is performed at the patient level, the maximum aggregation construction also provides a marginal coverage lower bound of $1-\alpha$ for each individual voxel.

\paragraph{Quantile aggregation.}

As an alternative to maximum aggregation and its requirement of simultaneous coverage of all voxels, we also propose a quantile-based patient-level nonconformity score that targets coverage of a prescribed fraction of voxels.
For a prescribed voxel coverage fraction $\delta \in (0, 1)$, the patient-level nonconformity score is defined as the empirical $\delta$-quantile of the voxelwise nonconformity scores.
More precisely, for a patient with input $\vb{x}$ and ground truth dose map $\vb{y}$, we define the nonconformity score as
\begin{equation}
\label{eq:patient-level-score-quantile}
s_{\delta}(\vb{x}, \vb{y}) \coloneq
\operatorname{Quantile}_{\delta}
\left(
\left\{
\abs{y_v - f_{\hat{\boldsymbol{\theta}}}(\vb{x})_v}
\right\}_{v=1}^V
\right).
\end{equation}
Each calibration patient therefore contributes the empirical $\delta$-quantile of its voxelwise residuals as a single scalar nonconformity score.
Here, the empirical quantile is defined as the $\lceil \delta V \rceil$-th smallest voxelwise nonconformity score.
After computing $\hat{q}$ from the patient-level calibration scores using~\eqref{eq:q-hat}, the resulting conformal prediction set for an input $\vb{x}$ is
\begin{equation}
\label{eq:conformal-prediction-set-quantile}
\mathcal{C}_{\delta}(\vb{x}) =
\left\{
\vb{y} \in \mathcal{Y} :
\operatorname{Quantile}_{\delta}
\left(
\left\{
\abs{y_v - f_{\hat{\boldsymbol{\theta}}}(\vb{   x})_v}
\right\}_{v=1}^V
\right)
\leq \hat{q}
\right\}.
\end{equation}
Equivalently, $\mathcal{C}_{\delta}(\vb{x})$ contains all candidate dose maps $\vb{y}$ for which the voxelwise dose values $y_v$ lie within their respective voxelwise prediction intervals $\mathcal{I}_v(\vb{x}) \coloneq [f_{\hat{\boldsymbol{\theta}}}(\vb{x})_v -\hat{q}, \,
f_{\hat{\boldsymbol{\theta}}}(\vb{x})_v +\hat{q}]$ for at least a fraction $\delta$ of the voxels.
This approach thus provides a guarantee on the voxel coverage fraction
\begin{equation}
C(\vb{x}, \vb{y}) \coloneq \frac{1}{V} \sum_{v=1}^V
\mathbf{1}_{\mathcal{I}_v(\vb{x})}(y_v),
\end{equation}
where $\mathbf{1}_{\mathcal{I}_v(\vb{x})}(y_v)$ equals one if $y_v \in \mathcal{I}_v(\vb{x})$ and zero otherwise.
Under exchangeability of the calibration data and a future test pair $(\vb{X}^{(\text{test})}, \vb{Y}^{(\text{test})})$, it follows from~\eqref{eq:marginal-coverage-guarantee} that this construction satisfies
\begin{equation}
\label{eq:quantile-nc-score-guarantee}
\mathbb{P}\left[ 
C(\vb{X}^{(\text{test})}, \vb{Y}^{(\text{test})})
\geq \delta
\right] \geq 1 - \alpha.
\end{equation}
That is, this construction guarantees with probability at least $1 - \alpha$ in the sense of~\eqref{eq:marginal-coverage-guarantee} that the voxelwise prediction intervals of a future patient contain their respective true dose values for at least the prescribed minimum fraction $\delta$ of its voxels.

Similar coverage events have also been considered in previous work.
In dose calculation, \citet{fisher-2024} used RCPS to calibrate voxelwise quantile regression intervals so that,
with a probability of at least the prescribed level over the calibration data,
the expected voxel miscoverage on future data does not exceed a prescribed level.
In contrast, our split CP quantile aggregation construction guarantees,
with a probability of at least the user-defined level,
that a future patient attains at least a prescribed minimum level of voxel coverage.
The constructions also differ:
\citet{fisher-2024} used RCPS with a voxel miscoverage risk,
while we use split CP with a patient-level nonconformity score obtained by quantile aggregation of voxelwise scores.
In image segmentation,
\citet{mossina-2024} used CRC with a thresholded binary loss based on whether a prescribed fraction of ground truth labels are covered by their corresponding pixelwise label prediction sets,
while~\citet{mossina-friedrich-2025} used split CP with morphological dilation to target coverage of a prescribed fraction of the ground truth segmentation mask.
Although these approaches target analogous coverage events and provide analogous guarantees for image segmentation, 
their constructions and the resulting prediction sets differ substantially from ours,
as discussed above for maximum aggregation.

\subsection{Groupwise calibration for region-specific guarantees}

In radiotherapy dose prediction, it may be desirable to control coverage separately within distinct anatomical regions rather than over the dose map as a whole.
In particular, for the quantile aggregation approach, attaining the prescribed voxel coverage fraction globally does not imply that the same coverage level is attained locally within each anatomical region.
For example, a method could exhibit systematically insufficient coverage in one anatomical region while still attaining the prescribed global coverage fraction through higher coverage elsewhere.
This is particularly relevant in this context as the clinical consequences of a prediction error can depend on the anatomical region in which it is committed.

A common approach to obtain region-specific coverage guarantees is groupwise calibration~\citep{angelopoulos-2023},
which we can apply to the patient-level constructions introduced above.
Specifically, we partition the voxels of each patient into predefined regions of interest indexed by $g \in \mathcal{G}$
(e.g. anatomically or clinically defined regions such as target volumes or organs at risk)
and compute the patient-level nonconformity score separately within each group, using only the voxelwise nonconformity scores belonging to that group.
Calibration is then performed separately for each group, yielding a group-specific threshold $\hat{q}_g$.
For each group $g \in \mathcal{G}$, the groupwise threshold $\hat{q}_g$ is then used to define a groupwise conformal prediction set $\mathcal{C}_g(\vb{x})$ for the dose values of the voxels within region $g$,
thus providing a patient-level coverage guarantee specific to that region.

The motivation for groupwise calibration differs slightly for the two patient-level constructions.
For maximum aggregation, even with standard global calibration, the patient-level coverage guarantee already implies coverage of all voxels within any individual anatomical region~\eqref{eq:max-score-induced-voxelwise-guarantee},
but it uses a single threshold to satisfy the stronger requirement of covering the true dose of all voxels across the whole patient.
Groupwise calibration instead allows each anatomical region to be calibrated separately,
thus yielding region-specific thresholds that allow prediction interval widths to differ across regions
while providing a guarantee of simultaneous coverage of all voxels within each region.
For quantile aggregation, groupwise calibration additionally ensures that the prescribed minimum level of voxel coverage is attained within each region separately, which is not implied by the global guarantee.

\subsection{Adaptive intervals using model uncertainty estimates}

Using the voxelwise nonconformity score in~\eqref{eq:voxelwise-residual}, the constructions introduced above yield prediction intervals whose width is fixed globally or within each calibrated group.
However, the reliability of the model predictions may vary greatly between patients, and even across voxels within the same patient.
Therefore, it is desirable for the prediction intervals to be adaptive:
their widths should vary according to the local reliability of the model predictions.

Predictive uncertainty provides a natural measure of prediction reliability that can be used to make the conformal intervals adaptive.
A common approach for conformal regression is to incorporate a heuristic measure of uncertainty as a scaling factor into the nonconformity score in order to obtain adaptive prediction intervals~\citep{angelopoulos-2023}.
Here, we use the predictive uncertainty estimated by a deep ensemble as this local scaling factor.
Intuitively, an inaccurate prediction associated with low uncertainty should be treated as more nonconforming than an equally inaccurate prediction associated with high uncertainty.
The voxelwise nonconformity score in~\eqref{eq:voxelwise-residual} thus becomes
\begin{equation}
\label{eq:uncertainty-scaled-nc-score}
s_v(\vb{x}, y_v) \coloneq
\frac{\abs{y_v - f_{\hat{\boldsymbol{\theta}}}(\vb{x})_v}}{\hat{u}(\vb{x})_v},
\end{equation}
where $\hat{u}(\vb{x})_v$ denotes an estimate of the model's predictive uncertainty at voxel $v$.

The patient-level maximum and quantile nonconformity scores are then obtained by replacing the absolute error term $\abs{y_v - f_{\hat{\boldsymbol{\theta}}}(\vb{x})_v}$ in~\eqref{eq:patient-level-score-max} and~\eqref{eq:patient-level-score-quantile} with the
uncertainty-normalised term $\abs{y_v - f_{\hat{\boldsymbol{\theta}}}(\vb{x})_v} / \hat{u}(\vb{x})_v$
from~\eqref{eq:uncertainty-scaled-nc-score}.
After computing $\hat{q}$ from the patient-level calibration scores using~\eqref{eq:q-hat}, the corresponding conformal prediction sets are obtained from~\eqref{eq:conformal-prediction-set-max} and~\eqref{eq:conformal-prediction-set-quantile} respectively by making the same replacement.
This substitution can be applied to both global and groupwise constructions.

For a calibrated threshold $\hat{q}$, this approach yields the following voxelwise prediction interval:
\begin{equation}
\mathcal{I}_v(\vb{x}) \coloneq
[f_{\hat{\boldsymbol{\theta}}}(\vb{x})_v - \hat{u}(\vb{x})_v \, \hat{q}, \,
f_{\hat{\boldsymbol{\theta}}}(\vb{x})_v + \hat{u}(\vb{x})_v \, \hat{q}].
\end{equation}
Thus, the prediction interval at voxel $v$ has width $2 \hat{u}(\vb{x})_v \hat{q}$ and is therefore adaptive:
it grows wider where the estimated predictive uncertainty is greater and becomes narrower where the model is more confident.
Importantly, this modification does not affect the corresponding conformal coverage guarantees, which remain the same as for the unscaled constructions.

\subsection{Experiments and evaluation protocol}

DL-based radiotherapy dose prediction is generally formulated as a supervised regression task.
The input is typically composed of medical images along with anatomical contours identifying the principal regions of interest, such as planning target volumes (PTVs) and organs at risk (OARs), while the target output corresponds to a 3D dose map.
In this study, we use the publicly available head-and-neck (H\&N) cancer patient dataset from the OpenKBP --- AAPM Grand Challenge~\citep{openkbp-2021}.
This dataset is composed of 340 patients, divided into 200 training cases, 40 validation cases, and 100 test cases.
As a publicly available dataset of limited size, OpenKBP provides a relevant setting for evaluating UQ methods under data constraints representative of those commonly encountered in medical applications.
For each patient, a normalised planning CT image $\vb{x}_{\text{CT}} \in [0, 1]^{128 \times 128 \times 128}$ is provided along with binary masks identifying the patient body, up to seven OARs, and up to three PTVs
(the 70~Gy PTV is present for all patients, whereas the 63~Gy and 56~Gy PTVs are present only in some cases).
These inputs are combined into a single tensor of 12 channels, with channels corresponding to absent structures filled with zeroes.
Each patient is also associated with a reference dose distribution $\vb{y} \in \mathbb{R}_{\geq 0}^{128 \times 128 \times 128}$ which constitutes the prediction target.

\begin{figure}[!htb]
    \centering

    \begin{subfigure}{0.95 \textwidth}
    \centering
    \includegraphics[width=\linewidth]{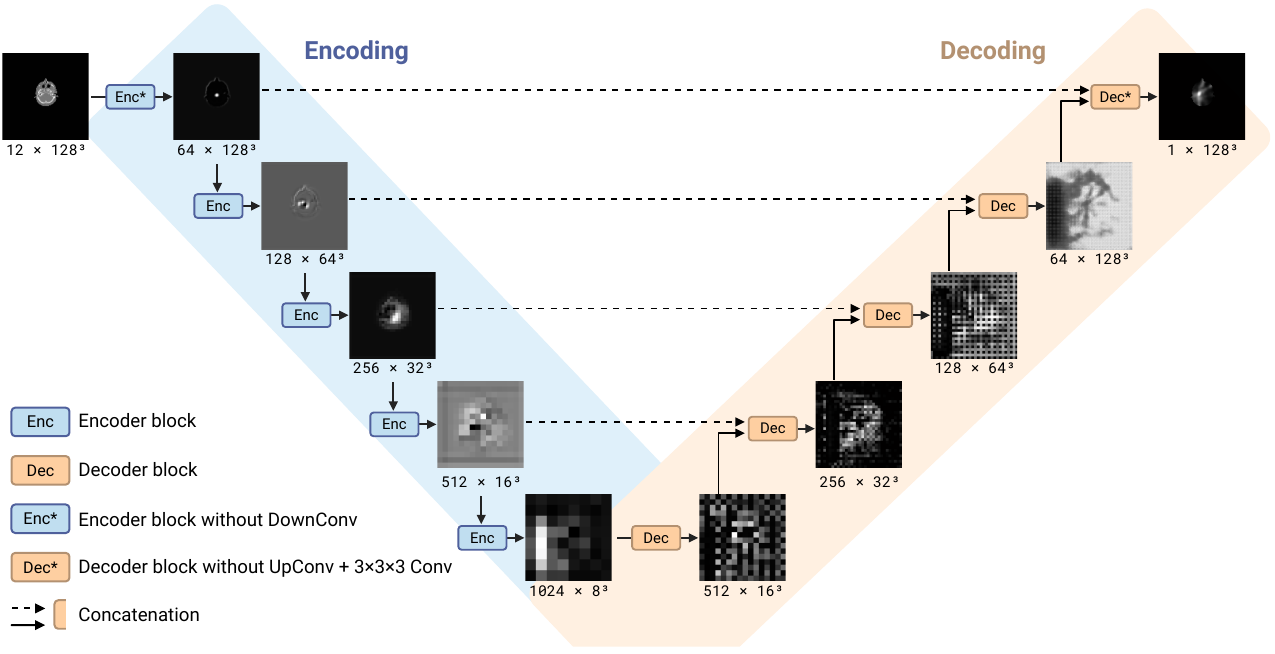}
    \caption{Architecture of the designed U-Net. A randomly selected feature map is shown at each level to illustrate how the network progressively transforms the input into a dose map.}
    \end{subfigure}

    \begin{subfigure}{0.95 \textwidth}
    \centering
    \includegraphics[width=\linewidth]{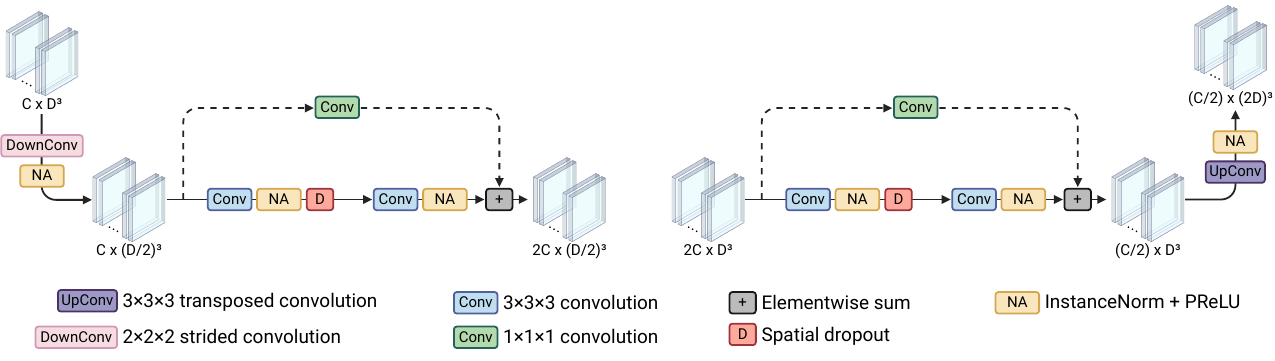}
    \caption{(Left) U-Net encoder block. (Right) U-Net decoder block.}
    \end{subfigure}

    \caption{Architecture of the 3D U-Net used for dose prediction.}
    \label{fig:dose-prediction-model}
\end{figure}

Dose prediction was performed using a 3D U-Net architecture~\citep{ronneberger-2015},
illustrated in Figure~\ref{fig:dose-prediction-model}.
The model was trained to minimise the mean absolute error (MAE) between the predicted and reference dose distributions within the patient body
using AdamW~\citep{loshchilov-adamw-2019} with an initial learning rate of $3 \times 10^{-4}$, a weight decay of $10^{-4}$, and a batch size of one.
During training,
spatial dropout~\citep{tompson-spatial-dropout-2015} was applied at the three deepest levels of the network, with dropout rates increasing with network depth ($p=0.05,0.10,0.15$).
Data augmentations consisted of random left-right flipping ($p=0.5$) and random affine transformations in the axial plane ($p=0.7$), consisting of rotations of up to $\pm 0.05 \ \text{rad}$ and translations of up to $\pm 8$ voxels along both axes of said plane.
The learning rate was reduced by half when the average validation loss did not improve for 20 epochs, with a cooldown of 10 epochs following each reduction.
The model parameters leading to the lowest average validation loss were retained for evaluation.

To obtain estimates of the model's predictive uncertainty, we used a deep ensemble of $M=16$ independently initialised and trained instances of the dose prediction model described above, all trained on the same dataset.
The variability across ensemble predictions is then used as a proxy for the epistemic component of predictive uncertainty,
which comes from our limited knowledge of the underlying data-generating process.
For an input $\vb{x}$, the final dose prediction was defined as the voxelwise mean
\begin{equation}
\label{eq:ensemble-mean-prediction}
\bar{\vb{y}} \coloneq \frac{1}{M} \sum_{m=1}^M f_{\hat{\boldsymbol{\theta}}_m}(\vb{x}),
\end{equation}
where $f_{\hat{\boldsymbol{\theta}}_m}(\vb{x})$ is the prediction made by the $m$-th model of the deep ensemble.
The predictive uncertainty was estimated by the voxelwise standard deviation across ensemble predictions:
\begin{equation}
\label{eq:deep-ensemble-uncertainty-estimate}
\hat{u}(\vb{x}) \coloneq \sqrt{
\frac{1}{M-1} \sum_{m=1}^M
\left(
f_{\hat{\boldsymbol{\theta}}_m}(\vb{x}) - \bar{\vb{y}}
\right)
\odot
\left(
f_{\hat{\boldsymbol{\theta}}_m}(\vb{x}) - \bar{\vb{y}}
\right)
},
\end{equation}
where $\odot$ is the Hadamard product and the square root is applied elementwise.
During evaluation, the mean dose predictions of the ensemble and the bounds of the voxelwise prediction intervals were clipped to the range $[0, 80]$~Gy.

The different split CP constructions introduced above were evaluated using the trained model ensemble and the held out test set of 100 patients provided by OpenKBP.
For each experiment, the 100 held out patients were randomly partitioned into a calibration set of $n_\text{cal}=50$ patients, on which the conformal methods were calibrated, and a test set of $50$ patients, on which they were evaluated.
Because the coverage guarantee in~\eqref{eq:marginal-coverage-guarantee} holds marginally over the calibration data and future test cases, the actual coverage obtained for any particular calibration set may vary.
To account for this variability, each experiment was repeated across $25$ random partitions of the held out data into calibration and test sets.
To allow for proper comparison, the same calibration-test partitions are used across all conformal methods.

For each calibration patient, voxelwise nonconformity scores were computed for the voxels within the patient's body using the ensemble mean prediction from~\eqref{eq:ensemble-mean-prediction}.
Throughout these experiments, only voxels within the patient's body were considered.
The voxelwise scores were then aggregated at the patient level using either the maximum or quantile construction introduced above.
For the quantile construction, the prescribed minimum voxel coverage fraction was set to $\delta=0.9$.
The nonconformity thresholds $\hat{q}$ were then computed from the patient-level calibration scores using~\eqref{eq:q-hat}, with a nominal coverage level of $1-\alpha = 0.9$.

For groupwise calibration, every voxel within a patient's body was sorted into one of three groups
\begin{equation}
g \in \mathcal{G} = \{ \text{PTV}, \text{OAR}, \text{Rest} \},
\end{equation}
where "Rest" included all body voxels belonging to neither a PTV nor an OAR.
For each new input $\vb{x}$, a groupwise conformal prediction set $\mathcal{C}_g(\vb{x})$ was then constructed for each region $g$ using its corresponding calibrated threshold $\hat{q}_g$.
Each of these conformal prediction sets therefore satisfies a distinct patient-level coverage guarantee specific to the corresponding anatomical region.
For the uncertainty-adaptive constructions, the voxelwise predictive uncertainty estimates $\hat{u}(\vb{x})_v$ derived from the deep ensemble using~\eqref{eq:deep-ensemble-uncertainty-estimate} were used to scale the voxelwise nonconformity scores, as described in~\eqref{eq:uncertainty-scaled-nc-score}.
The maximum and quantile constructions were evaluated for both global and groupwise calibration, using both standard and uncertainty-scaled voxelwise scores.

The different methods were first evaluated in terms of validity by measuring the empirical frequency with which their respective conformal coverage guarantees were satisfied on the test patients.
While the marginal coverage guarantee in~\eqref{eq:marginal-coverage-guarantee} ensures validity, it does not by itself ensure that the resulting prediction set is informative.
Indeed, arbitrarily large prediction sets can trivially achieve high coverage.
Therefore, we also evaluated the efficiency of the methods through the median width of the voxelwise prediction intervals for each patient, and examined how interval width varied across anatomical regions.
More precisely, for each random data split, we computed the mean across test patients of each patient's median voxelwise interval width.
Narrower intervals indicate greater efficiency when the desired coverage requirement is satisfied.
Otherwise, arbitrarily narrow intervals can be obtained through undercoverage.
Empirical coverage and interval width must therefore be considered jointly when assessing the practical usefulness of a method.
Also, empirical voxelwise coverage was evaluated within each anatomical region.
More precisely, for each random data split, we computed the mean across test patients of each patient's voxelwise coverage within all anatomical regions.
This was done because a method could in principle attain satisfactory global coverage despite systematic undercoverage in a specific subregion by compensating through higher coverage elsewhere.
The reported values correspond to the mean and standard deviation of these measures across the random data splits.

\section{Results}
\label{sec:results}

\begin{figure}[!htbp]
    \centering

    \begin{subfigure}{0.90 \textwidth}
    \centering
    \includegraphics[width=\linewidth]{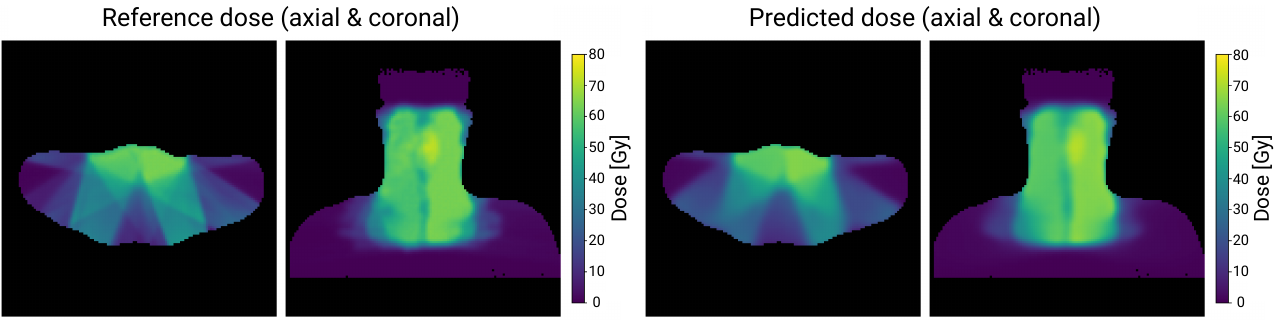}
    \caption{Reference and predicted dose distributions in axial and coronal views.}
    \end{subfigure}

    \begin{subfigure}{0.90 \textwidth}
    \centering
    \includegraphics[width=\linewidth]{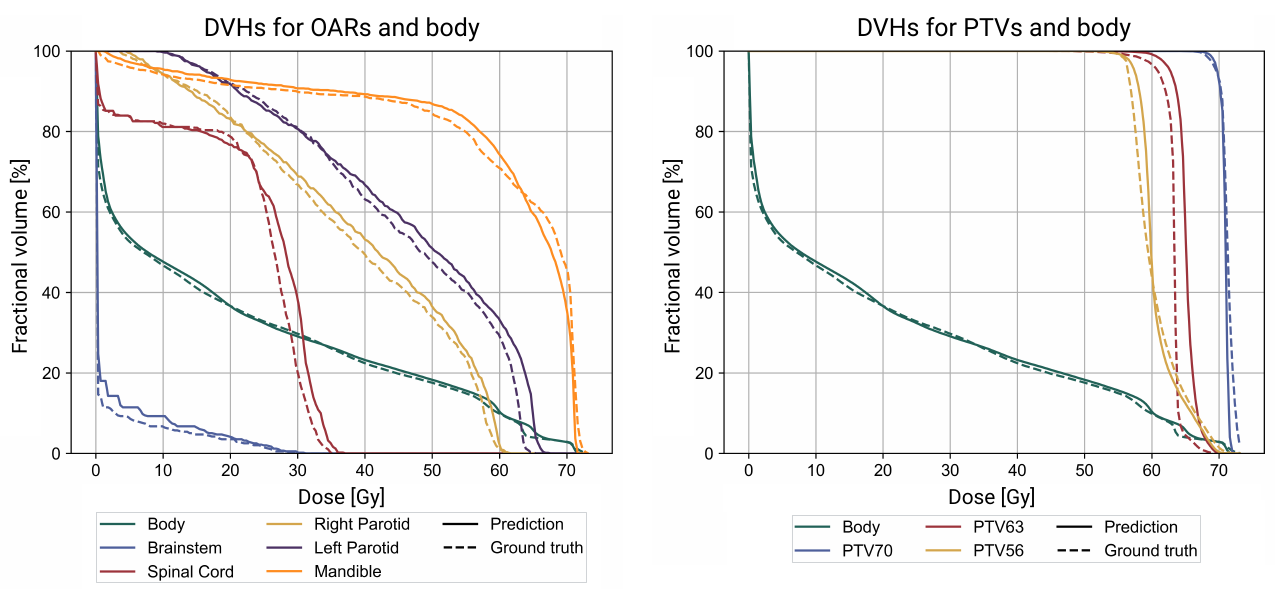}
    \caption{Dose-volume histograms (DVHs) for the OARs, PTVs and body.}
    \end{subfigure}

    \caption{Illustrative example of dose prediction by the deep ensemble for a test patient, shown alongside the reference dose and corresponding DVHs.}
    \label{fig:dose-prediction-example}
\end{figure}

The trained ensemble model achieved a mean body MAE of $2.5 \pm 1.1$~Gy over the 100 patients held out for evaluation.
Figure~\ref{fig:dose-prediction-example} shows the mean dose prediction from the ensemble alongside the corresponding reference dose distribution and dose-volume histograms (DVHs) for an example test patient (body MAE of $2.22$~Gy).

Tables~\ref{tab:max-global} and~\ref{tab:max-groupwise} present the results for the maximum aggregation construction defined in~\eqref{eq:patient-level-score-max} with global calibration and groupwise calibration respectively.
As expected, the empirical coverages of the conformal prediction set $\mathcal{C}_\text{max}$ and the group-specific conformal prediction sets $\mathcal{C}_\text{max}, g$
were close to the nominal coverage level $1 -\alpha =0.9$
for both the standard voxelwise nonconformity score and the uncertainty-scaled score.
Empirical voxelwise coverage was essentially $100\%$ across all anatomical regions for both calibration strategies and both scores.
As further discussed in Section~\ref{sec:discussion}, this suggests substantial conservativeness.
With global calibration, the median width of the voxelwise prediction intervals was greater across all anatomical regions with the uncertainty-scaled score,
reaching the maximum possible width of $80$~Gy in the OARs (due to clipping of the dose interval).
With groupwise calibration, the median width of the voxelwise prediction intervals was substantially narrower within the OAR and PTV regions than for global calibration,
while staying essentially the same within the Rest region for the standard score and decreasing for the uncertainty-scaled score.
The median voxelwise interval widths were also greater with the uncertainty-scaled score in the OAR and Rest regions, while a slight reduction was observed in the PTV compared to the standard score.

\begin{table}[ht]
\centering
\small
\setlength{\tabcolsep}{4pt}

\caption{
\textbf{Maximum aggregation, global calibration.}
Results for the construction defined in~\eqref{eq:patient-level-score-max} with a nominal coverage level of $1-\alpha=0.9$.
Values are reported as mean (standard deviation) across 25 random data splits (50 calibration and 50 test patients).
}
\label{tab:max-global}

\begin{tabular}{@{}l c ccc ccc@{}}
\toprule
\multirow{3}{*}{\makecell[l]{Voxelwise\\nonconformity\\ score}}
&
\multirow{2}{*}{\makecell{Empirical coverage\\of $\mathcal{C}_{\max}$ [\%]}}
&
\multicolumn{3}{c}{\makecell{Empirical voxelwise\\coverage [\%]}}
&
\multicolumn{3}{c}{\makecell{Median width of voxelwise\\prediction intervals [Gy]}}
\\
\cmidrule(lr){3-5}
\cmidrule(lr){6-8}
&
&
OAR & PTV & Rest
&
OAR & PTV & Rest
\\

\midrule

Standard~\eqref{eq:voxelwise-residual}
& $91\,(5)$
& $100\,(0)$
& $100\,(0)$
& $100\,(0)$
& $56.1\,(0.9)$
& $45.5\,(1.2)$
& $43.9\,(1.1)$
\\

\addlinespace

Scaled~\eqref{eq:uncertainty-scaled-nc-score}
& $90\,(5)$
& $100\,(0)$
& $100\,(0)$
& $99.97\,(0.02)$
& $80\,(0)$
& $50.4\,(3)$
& $69\,(7)$
\\

\bottomrule
\end{tabular}
\end{table}

\begin{table}[ht]
\centering
\small
\setlength{\tabcolsep}{3.5pt}
\caption{
\textbf{Maximum aggregation, groupwise calibration.}
Results for the construction defined in~\eqref{eq:patient-level-score-max} with a nominal coverage level of $1-\alpha=0.9$ for each group.
Values are reported as mean (standard deviation) across 25 random data splits (50 calibration and 50 test patients).
}
\label{tab:max-groupwise}

\begin{tabular}{@{}l ccc ccc ccc@{}}
\toprule
\multirow{3}{*}{\makecell[l]{Voxelwise\\nonconformity\\score}}
&
\multicolumn{3}{c}{\makecell{Empirical coverage\\of $\mathcal{C}_{\max, g}$ [\%]}}
&
\multicolumn{3}{c}{\makecell{Empirical voxelwise\\coverage [\%]}}
&
\multicolumn{3}{c}{\makecell{Median width of voxelwise\\prediction intervals [Gy]}}
\\
\cmidrule(lr){2-4}
\cmidrule(lr){5-7}
\cmidrule(lr){8-10}
&
OAR & PTV & Rest
&
OAR & PTV & Rest
&
OAR & PTV & Rest
\\

\midrule

Standard~\eqref{eq:voxelwise-residual}
&
$90\,(5)$ & $89\,(5)$ & $91\,(5)$
&
$99.6\,(0.3)$ & $99.5\,(0.3)$ & $99.98\,(0.02)$
&
$36.5\,(1.1)$ & $24.8\,(0.8)$ & $43.9\,(1.1)$ \\

\midrule

\makecell[l]{Scaled~\eqref{eq:uncertainty-scaled-nc-score}}
&
$90\,(5)$ & $89\,(6)$ & $90\,(5)$
&
$99.92\,(0.07)$ & $99.89\,(0.08)$ & $99.97\,(0.02)$
&
$40\,(5)$ & $23.3\,(0.7)$ & $55\,(3)$ \\

\bottomrule
\end{tabular}
\end{table}

Tables~\ref{tab:quantile-global} and~\ref{tab:quantile-groupwise} present the results for the quantile aggregation construction defined in~\eqref{eq:patient-level-score-quantile} with  $\delta=0.9$ using global calibration and groupwise calibration respectively.
As expected, the empirical coverages of the conformal prediction set $\mathcal{C}_\delta$ and the group-specific conformal prediction sets $\mathcal{C}_{\delta, g}$ were close to the nominal coverage level $1-\alpha=0.9$ for both calibration strategies and both voxelwise nonconformity scores.
With global calibration, empirical voxelwise coverage was particularly high in the PTV region with the standard score,
whereas groupwise calibration yielded more similar voxelwise coverage across anatomical regions.
With groupwise calibration, the median width of the voxelwise prediction intervals in the OAR and PTV regions was narrower than for global calibration,
particularly in the PTV
where the median width decreased from $17.9$~Gy to $8.6$~Gy with the standard score.
Uncertainty scaling reduced median interval widths across all anatomical regions with global calibration;
however, with groupwise calibration,
uncertainty scaling had little effect on the median interval width in the OAR and PTV regions,
while reducing it from $19.2$~Gy to $12.6$~Gy in the rest of the body.

\begin{table}[ht]
\centering
\small
\setlength{\tabcolsep}{4pt}

\caption{
\textbf{Quantile aggregation, global calibration.}
Results for the construction defined in~\eqref{eq:patient-level-score-quantile} with $\delta=0.9$ and a nominal coverage level of $1-\alpha=0.9$.
Values are reported as mean (standard deviation) across 25 random data splits (50 calibration and 50 test patients).
}
\label{tab:quantile-global}

\begin{tabular}{@{}l c ccc ccc@{}}
\toprule
\multirow{3}{*}{\makecell[l]{Voxelwise\\nonconformity\\ score}}
&
\multirow{2}{*}{\makecell{Empirical coverage\\of $\mathcal{C}_{\delta}$ [\%]}}
&
\multicolumn{3}{c}{\makecell{Empirical voxelwise\\coverage [\%]}}
&
\multicolumn{3}{c}{\makecell{Median width of voxelwise\\prediction intervals [Gy]}}
\\
\cmidrule(lr){3-5}
\cmidrule(lr){6-8}
&
&
OAR & PTV & Rest
&
OAR & PTV & Rest
\\

\midrule

Standard~\eqref{eq:voxelwise-residual}
& $90\,(4)$
& $96.0\,(1.7)$ & $99.1\,(0.5)$ & $93.5\,(1.6)$
& $17.9\,(1.3)$ & $17.9\,(1.3)$ & $17.7\,(1.1)$ \\

\midrule

\makecell[l]{Scaled~\eqref{eq:uncertainty-scaled-nc-score}}
& $90\,(6)$
& $96.0\,(1.1)$ & $95.5\,(1.4)$ & $93.5\,(1.2)$
& $15.5\,(0.6)$ & $8.6\,(0.3)$ & $12.2\,(1.0)$ \\

\bottomrule
\end{tabular}
\end{table}

\begin{table}[ht]
\centering
\small
\setlength{\tabcolsep}{4pt}
\caption{
\textbf{Quantile aggregation, groupwise calibration.}
Results for the construction defined in~\eqref{eq:patient-level-score-quantile} with $\delta=0.9$ and a nominal coverage level of $1-\alpha=0.9$ for each group.
Values are reported as mean (standard deviation) across 25 random data splits (50 calibration and 50 test patients).
}
\label{tab:quantile-groupwise}

\begin{tabular}{@{}l ccc ccc ccc@{}}
\toprule
\multirow{3}{*}{\makecell[l]{Voxelwise\\nonconformity\\ score}}
&
\multicolumn{3}{c}{\makecell{Empirical coverage\\of $\mathcal{C}_{\delta, g}$ [\%]}}
&
\multicolumn{3}{c}{\makecell{Empirical voxelwise\\coverage [\%]}}
&
\multicolumn{3}{c}{\makecell{Median width of voxelwise\\prediction intervals [Gy]}}
\\
\cmidrule(lr){2-4}
\cmidrule(lr){5-7}
\cmidrule(lr){8-10}
&
OAR & PTV & Rest
&
OAR & PTV & Rest
&
OAR & PTV & Rest
\\

\midrule

Standard~\eqref{eq:voxelwise-residual}
&
$91\,(6)$ & $90\,(6)$ & $90\,(4)$
&
$94.8\,(1.1)$ & $95.3\,(1.2)$ & $95.1\,(1.1)$
&
$15.4\,(0.6)$ & $8.6\,(0.3)$ & $19.2\,(0.7)$ \\

\midrule

\makecell[l]{Scaled~\eqref{eq:uncertainty-scaled-nc-score}}
&
$90\,(5)$ & $89\,(7)$ & $90\,(5)$
&
$94.5\,(1.3)$ & $94\,(2)$ & $94.4\,(1.2)$
&
$14.6\,(0.7)$ & $8.1\,(0.6)$ & $12.6\,(1.1)$ \\

\bottomrule
\end{tabular}
\end{table}

\begin{figure}[!htbp]
    \centering
    \includegraphics[width=\textwidth]{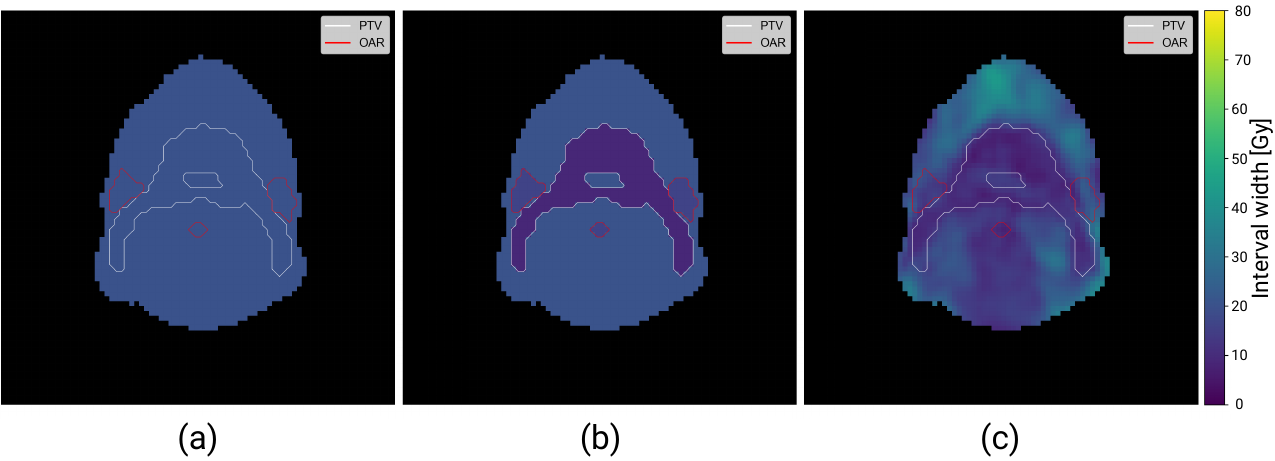}
    \caption{
    Width of the voxelwise prediction intervals for a test patient using quantile aggregation ($\delta=0.9$).
    This representation highlights the effects of groupwise calibration and uncertainty scaling on adaptivity.
    (a)~Standard voxelwise nonconformity score with global calibration.
    (b)~Standard voxelwise nonconformity score with groupwise calibration.
    (c)~Uncertainty-scaled voxelwise nonconformity score with groupwise calibration.
    }
    \label{fig:adaptivity}
\end{figure}

Figure~\ref{fig:adaptivity} shows how voxelwise prediction interval width varies across the predicted dose map for a random test patient using different quantile aggregation approaches.
With the standard voxelwise nonconformity score and global calibration,
voxelwise prediction intervals were essentially constant in width across voxels and patients,
with any variation arising solely from post-processing clipping.
Groupwise calibration led to region-dependent interval widths, with a different width for each anatomical group but remaining essentially constant across the voxels of that group and across patients.
In contrast, uncertainty scaling produced locally adaptive voxelwise prediction intervals, with widths varying both from voxel to voxel and from patient to patient
according to the model's estimated predictive uncertainty.

\section{Discussion}
\label{sec:discussion}

\subsection{Patient-level nonconformity scores}
\label{sec:discussion-patient-level-nc-scores}

Our maximum aggregation construction defined in~\eqref{eq:patient-level-score-max} provides a particularly strong form of coverage control,
guaranteeing simultaneous coverage of all considered voxels of a future patient with probability at least $1-\alpha$,
as stated in~\eqref{eq:max-nc-score-guarantee}.
Empirically,
as expected from the split CP marginal coverage guarantee,
coverage of the conformal prediction sets remained close to the intended level for both the global and groupwise constructions (Tables~\ref{tab:max-global} and~\ref{tab:max-groupwise}).
However, requiring that a common threshold ensure simultaneous coverage of the true dose at all voxels of a future patient is highly demanding for a high-dimensional output such as a 3D dose map.
Because the patient-level nonconformity score is determined by the single most nonconforming voxel,
a single aberrant voxel among the large number of voxels in the patient's body suffices to substantially increase the threshold needed for coverage, thus widening the voxelwise prediction intervals across the entire dose map. 
The large interval widths observed for both global and groupwise calibration are consistent with this expected behaviour.
We showed in~\eqref{eq:max-score-induced-voxelwise-guarantee} that the patient-level marginal coverage guarantee of the maximum construction induces a marginal coverage lower bound of $1-\alpha$ for the prediction interval of any fixed voxel,
despite split CP being performed at the patient level.
This property is theoretically appealing because it recovers a voxel-specific marginal guarantee from patient-level calibration.
However, the empirical voxelwise coverage was close to $100\%$ for both global and groupwise calibration,
reflecting the conservative nature of the maximum construction in this high-dimensional setting.

Our quantile aggregation construction defined in~\eqref{eq:patient-level-score-quantile} guarantees with probability at least $1-\alpha$ that a future patient attains a prescribed minimum level of voxel coverage,
as stated in~\eqref{eq:quantile-nc-score-guarantee}.
Empirically,
as expected from the split CP marginal coverage guarantee,
coverage of the conformal prediction sets remained close to the intended level for both the global and groupwise constructions (Tables~\ref{tab:quantile-global} and~\ref{tab:quantile-groupwise}).
By basing the patient-level nonconformity score on a prescribed quantile of the voxelwise nonconformity scores,
isolated extreme voxelwise errors do not necessarily affect the conformal threshold.
This makes the construction particularly well suited to high-dimensional outputs such as dose maps,
where a few large local errors need not dictate the prediction interval widths across the many remaining voxels.
The resulting voxelwise prediction intervals were substantially narrower than those produced by maximum aggregation,
although this difference in interval width should not be interpreted as evidence of greater efficiency since the two constructions target different coverage events.
A potential concern with the globally calibrated quantile construction is that its coverage guarantee holds over the patient as a whole;
it does not imply any guarantee on the coverage level attained within a specific anatomical subregion.
In principle, insufficient coverage in one region could be compensated by higher coverage elsewhere such that the coverage requirement remains satisfied globally.
This potential concern is particularly relevant in dose prediction,
where systematic undercoverage of a clinically important region, such as OARs, could be masked by higher coverage elsewhere.
In our experiments, no such regional undercoverage was observed,
with empirical coverage of the voxelwise prediction intervals exceeding $90\%$ in all anatomical regions.
Notably, with the standard (unscaled) voxelwise nonconformity score and global calibration, voxelwise coverage was substantially higher in the PTV than in the OAR and Rest regions (Table~\ref{tab:quantile-global});
this disparity disappeared with uncertainty scaling and was not observed with groupwise calibration (Table~\ref{tab:quantile-groupwise}).

\subsection{Groupwise calibration}

Groupwise calibration provides region-specific coverage guarantees and allows the use of a distinct conformal threshold in each anatomical region.
Empirically, coverage of the group-specific conformal prediction sets remained close to the intended level for both maximum and quantile constructions,
suggesting as expected
that the corresponding region-specific coverage guarantees were satisfied in practice.
Groupwise calibration also led to different voxelwise prediction interval widths across anatomical regions, since each region was assigned its own conformal threshold.
For the maximum aggregation construction,
groupwise calibration substantially reduced interval widths in the OAR and PTV regions for both voxelwise nonconformity scores,
while leaving the median width in the Rest region essentially unchanged for the standard score but reducing it for the uncertainty-scaled score (Tables~\ref{tab:max-global} and~\ref{tab:max-groupwise}).
For the standard voxelwise score, this pattern is consistent with the maximum nonconformity score being driven primarily by extreme voxelwise nonconformity scores arising in the larger Rest region;
calibrating the regions separately prevents these extreme scores from widening the voxelwise prediction intervals in the OAR and PTV regions.
For quantile aggregation, groupwise calibration also affected the voxelwise prediction intervals differently across anatomical regions (Tables~\ref{tab:quantile-global} and~\ref{tab:quantile-groupwise}).
With the standard voxelwise nonconformity score, interval widths decreased in the OAR and more significantly in the PTV, whereas a slight increase was observed in the rest of the body.
This suggests that the globally calibrated threshold was relatively conservative for the PTV, while the Rest region required a slightly larger region-specific threshold.
Consistent with this interpretation, the relative overcoverage of the voxelwise prediction intervals observed in the PTV with global calibration disappeared with groupwise calibration.
These results show that groupwise calibration does not simply reduce interval widths uniformly;
rather, it allows split CP to reflect regional differences in the distribution of nonconformity scores.
This can substantially reduce unnecessary conservatism in some regions, while also providing region-specific coverage guarantees.
Using finer anatomical groups, such as individual organs rather than combining OARs, could further reflect regional differences.
Intermediate regions could also be considered to better characterise the behaviour of the method in areas with strong dose gradients.

\subsection{Adaptive intervals using model uncertainty estimates}

Although groupwise calibration allows the width of voxelwise prediction intervals to vary across anatomical regions, the standard voxelwise nonconformity score still produces intervals whose widths are essentially constant within each region and across patients.
However, prediction difficulty can vary substantially between patients and between voxels, even within the same anatomical region.
Ideally, the widths of voxelwise prediction intervals should adapt locally, growing wider where the model is more likely to make errors and narrower where its predictions are more reliable.
To introduce this local adaptivity, we scaled the voxelwise nonconformity score by the predictive uncertainty estimated from a deep ensemble.
The resulting voxelwise prediction intervals thus have widths proportional to the local uncertainty estimate, allowing them to vary across voxels and across patients.

As illustrated in Figure~\ref{fig:adaptivity}, the uncertainty-scaled voxelwise nonconformity score~\eqref{eq:uncertainty-scaled-nc-score} introduced local adaptivity:
interval width varied across voxels according to the estimated predictive uncertainty.
In contrast, the standard score resulted in a fixed interval width across patients, or fixed within each anatomical region, regardless of the model's predictive uncertainty.
This property is important regardless of its effect on median interval width, since locally adaptive intervals can reflect informative variations in prediction difficulty.

For quantile aggregation, uncertainty scaling reduced the median width of the voxelwise prediction intervals across all anatomical regions.
With global calibration, uncertainty scaling substantially reduced the median interval width in the PTV and Rest regions, with a smaller reduction in the OAR (Table~\ref{tab:quantile-global}).
In the PTV, where the largest reduction in interval width was observed, this reduction was also accompanied by the disappearance of the relative overcoverage of the voxelwise prediction intervals.
This suggests that uncertainty scaling adjusted interval width more appropriately across voxels, assigning narrower intervals to voxels associated with lower uncertainty and wider intervals for voxels where the model's predictive uncertainty was higher.

Interestingly, the width-reducing effects of groupwise calibration and uncertainty scaling were not additive in the OAR and PTV regions (Table~\ref{tab:quantile-groupwise}).
Indeed, when uncertainty scaling was applied to the groupwise quantile construction,
the median width of the voxelwise prediction intervals decreased only slightly in the OARs and PTVs,
while it was substantially reduced in the rest of the body.
This overlap may be explained by the fact that both approaches account for heterogeneity, but at different levels:
groupwise calibration accounts for systematic differences in error distributions between anatomical regions,
whereas uncertainty scaling accounts for local differences in prediction difficulty between individual voxels.
The latter provides a finer form of adaptivity,
but because the PTV and OAR regions are considerably homogeneous with respect to dose prediction,
much of the local adaptation achieved by uncertainty scaling can also be captured by groupwise calibration.
In contrast, the Rest region is much larger and encompasses a much more heterogeneous range of voxels and differing local contexts;
therefore, the local variation captured there by uncertainty scaling cannot be well represented by groupwise calibration alone.
Nevertheless, the two methods serve distinct purposes, and should not be regarded as redundant within OARs and PTVs.
Uncertainty scaling provides the local adaptivity required for the interval width to vary between voxels and patients according to estimated predictive uncertainty,
whereas groupwise calibration provides separate coverage control for distinct anatomical regions.

In contrast to quantile aggregation, uncertainty scaling did not improve the maximum aggregation construction in terms of prediction interval width;
instead, it generally led to wider intervals (Tables~\ref{tab:max-global} and~\ref{tab:max-groupwise}).
This is expected, as uncertainty scaling can exacerbate the issue of isolated extreme errors that was discussed in Section~\ref{sec:discussion-patient-level-nc-scores} for maximum aggregation.
Because the patient-level maximum is determined by a single voxel, a large prediction error combined with underestimated uncertainty in just one voxel of the patient is enough to produce an extreme nonconformity score, thus inflating the conformal threshold required for the whole volume.
Overall, these results suggest that uncertainty scaling is particularly beneficial when combined with quantile aggregation, where it introduces informative local adaptivity
without the excessive sensitivity to extreme scores that is inherent to the maximum construction.

\subsection{Clinical relevance and limitations}

The proposed split CP constructions provide an interpretable complement to dose predictions by assigning them voxelwise prediction intervals with explicit patient-level coverage guarantees.
In particular, our quantile aggregation construction provides control over the probability that a future patient attains a prescribed minimum level of voxel coverage,
while groupwise calibration extends this control to clinically relevant anatomical regions.
With uncertainty scaling, the resulting prediction intervals can adapt locally using the estimated predictive uncertainty, thus allowing them to become wider or narrower according to the estimated prediction reliability.
In contrast to typical UQ methods in dose prediction, the explicit coverage guarantees provided by split CP give the resulting prediction intervals a direct interpretation that raw uncertainty estimates alone cannot provide.
This framework could serve as a guide for physicians and dosimetrists using knowledge-based planning.
For example, uncertainty bounds could provide a quantitative basis for identifying cases that are too uncertain for reliable automated planning,
particularly those at risk of reaching dose tolerances of critical structures or underdosing target volumes.

However, the voxelwise prediction intervals remain wide.
While this width is informative about prediction reliability on a voxel-by-voxel scale
and reflects the stringent nature of the coverage events targeted in this work,
it may limit the clinical usefulness of the resulting voxelwise dose intervals.
Beyond prediction difficulty, this limitation may arise in part from the very nature of dose prediction:
the variability among clinically acceptable dose distributions represented in the data.
Indeed, a voxelwise deviation from the reference dose map does not necessarily represent a clinically meaningful prediction error.
For instance, voxelwise errors are amplified near steep dose gradients, where small spatial differences between otherwise similar dose maps can result in large local dose differences.
Although supervised learning typically considers a single reference plan,
multiple treatment plans corresponding to different tradeoffs between competing clinical objectives can in fact coexist for the same patient,
leading to substantially different local dose distributions that remain clinically acceptable~\citep{craft-2012}.
Variability among the reference plans may therefore contribute to wider voxelwise prediction intervals;
these wider intervals reflect uncertainty regarding the reproduction of the reference dose at each voxel
rather than whether a different dose distribution is clinically acceptable.
Using a more homogeneous set of treatment cases could reduce some of this variability and potentially lead to narrower prediction intervals.

Furthermore, if multiple clinically acceptable treatment plans exist for the same anatomy,
voxelwise prediction intervals can combine incompatible solutions.
For example, consider two clinically acceptable treatment plans for a head-and-neck patient for whom one parotid gland may be sacrificed to spare the other:
one plan may deliver a high dose to one parotid gland and a low dose to the other,
whereas the second plan may make the opposite trade-off.
Prediction intervals constructed at each voxel thus seek to cover these different possibilities locally,
by having the voxelwise prediction interval of each of these voxels span both low and high dose values.
Consequently, the resulting patient-level prediction set can also contain dose distributions that do not correspond to any clinically acceptable treatment plan,
such as a plan assigning a high dose to both parotid glands.
That is, combining the voxelwise prediction intervals into a patient-level prediction set
without accounting for dependencies across voxels
allows combinations of voxelwise dose values that do not correspond to any plausible dose distribution;
this may lead to unnecessarily large prediction sets, as these combinations could be excluded without excluding clinically acceptable dose distributions.
Future work on conformal dose prediction should therefore investigate approaches that can disentangle these different solutions;
this consideration also applies to related frameworks such as RCPS and CRC.

\section{Conclusion}
\label{sec:conclusion}

In this work, we investigated the application of split CP to DL-based radiotherapy dose prediction for treatment planning.
The main challenge in extending split CP to this task was the mismatch between the level at which exchangeability holds (patients) and the level at which prediction intervals are desired (voxels).
We addressed this issue by defining patient-level nonconformity scores that nonetheless yield voxelwise prediction intervals satisfying coverage guarantees formulated at the patient level.
We proposed two such patient-level constructions:
one based on maximum aggregation of voxelwise nonconformity scores, which guarantees simultaneous coverage of the true dose in all voxels of a future patient with probability at least the user-defined level and further implies a marginal coverage lower bound for each individual voxel,
and the other based on quantile aggregation of voxelwise scores,
which guarantees that a future patient attains at least a prescribed minimum level of voxel coverage with probability at least the user-defined level.
We further investigated groupwise calibration to obtain region-specific coverage guarantees and predictive uncertainty scaling to construct locally adaptive voxelwise prediction intervals.

Empirically, our proposed patient-level constructions achieved coverage consistent with the nominal coverage levels.
Maximum aggregation provided the strictest coverage guarantee but resulted in highly conservative prediction intervals,
illustrating the cost of requiring simultaneous coverage in a high-dimensional setting such as dose prediction.
On the other hand, quantile aggregation produced substantially narrower intervals
while guaranteeing, with a user-defined probability,
that a future patient attains at least a prescribed level of voxel coverage.
Groupwise calibration further provided region-specific coverage guarantees,
while uncertainty scaling enabled locally adaptive prediction intervals whose widths varied according to the model's estimated predictive uncertainty.

Overall, these results demonstrate that split CP can provide explicit guarantees for radiotherapy dose prediction,
while also highlighting the limitations of using voxel coverage as the target of conformal guarantees in this task.
The large interval widths observed may limit the practical usefulness of the resulting voxelwise prediction intervals; however, they remain informative about the difficulty of reproducing the reference dose at the voxelwise level.
More fundamentally, voxelwise coverage does not fully capture dependencies across voxels
or the existence of multiple clinically acceptable dose distributions for the same input anatomy.
Future work on conformal dose prediction should therefore investigate approaches
that better account for dependencies across voxelwise dose values when constructing patient-level prediction sets,
and target clinically meaningful patient-level properties of the dose distribution rather than voxel coverage alone.
This combination of UQ and formal guarantees makes split CP a promising framework for future research towards an eventual clinical integration of DL-based radiotherapy dose prediction.

\section*{Acknowledgements}
\label{sec:acknowledgements}

We acknowledge the support of the Natural Sciences and Engineering Research Council of Canada (NSERC), Discovery Grant No. 2024-04198.
R.B. acknowledges support from the Mitacs Accelerate program.

\section*{Data availability statement}
\label{sec:data-availability}

The data that support the findings of this study are openly available in the OpenKBP GitHub repository at \url{https://github.com/ababier/open-kbp}.

\bibliographystyle{plainnat}
\bibliography{references.bib}

\end{document}